\documentclass{czjphys}         
\def\lie{{\cal G}}
\def\eps{\epsilon}
\def\a{\alpha}

\def\d{\delta}
\def\l{\lambda}
\def\pr{\prime}
\def\br{\begin{eqnarray}}
\def\er{\end{eqnarray}}
\def\pa{\partial}
\def\o{\over}
\begin{document}
\title{T-Duality in  Affine NA  Toda Models}
\authori{{ J.F. Gomes},
 { G.M. Sotkov} and { A.H. Zimerman}}      
\addressi{Instituto de F\'\i sica Te\'orica - IFT/UNESP\\
Rua Pamplona 145\\
01405-900, S\~ao Paulo - SP, Brazil}
\authorii{}
\addressii{}
\authoriii{}    \addressiii{}
\authoriv{}     \addressiv{}
\authorv{}      \addressv{}
\authorvi{}     \addressvi{}
\headauthor{J.F. Gomes et al.}   
\headtitle{T-Duality in  Affine NA  Toda Models}
\lastevenhead{J.F. Gomes et al. T-Duality in  Affine NA  Toda Models}
\refnum{A}
\daterec{XXX}    
\issuenumber{0}  \year{2004}
\setcounter{page}{1}
\maketitle
\begin{abstract}
The   construction of Non Abelian affine Toda models 
is discussed  in terms of its underlying Lie algebraic structure.   
It is  shown that  a subclass of such non conformal two dimensional integrable models naturally 
leads to the construction of a pair of
actions which share the same spectra and are related by canonical transformations.
\end{abstract}

\section{Introduction}     

The affine Toda models consists of a class of relativistic  two dimensional 
integrable models admiting soliton solutions with non
trivial topological charge (e.g. the abelian affine Toda models).
Among such models we encounter certain   Non Abelian affine (NA) Toda models 
  admiting electrically charged solitons \cite{tau}.  In general,  the NA Toda models admit  solitons
with non trivial internal symmetry structure.  
 The formulation and classification of such  models with its global symmetry structure is given 
in terms of the decomposition of an underlying Lie algebraic structure 
according to a grading operator $Q$ and, in terms of a pair of constant generators 
$\epsilon_{\pm}$ of grade $\pm 1$. In particular, integrable perturbations of the WZW model characterized
 by $\epsilon_{\pm}$ describe the dynamics of fields parametrizing the zero grade subalgebra $\lie_0$.   
The action manifests  chiral symmetry associated to the subalgebra 
$\lie_0^0 \subset \lie_0$ due to the fact that $Y \in \lie_0^0, [Y, \epsilon_{\pm}]=0$.
The existence of such  subalgebra allows the implementation of 
subsidiary constraints within $\lie_0^0$ and   the reduction of  the model from the 
group $G_0$ to the coset $G_0/G_0^0$. The structure of the coset $G_0/G_0^0$  
viewed according to axial or vector
gauging  leads to different parametrizations and different actions, namely axial or vector actions.

We first discuss the general construction of NA Toda models 
in terms of the gauged WZW model. Next, we discuss the 
structure of the coset $G_0/G_0^0 = SL(2)\otimes U(1)^{n-1}/ U(1)$ and $G_0/G_0^0 = SL(3)/ SL(2)\otimes U(1)$
according to axial and vector
gaugings and explicitly construct the associated lagrangians. 
 Finally, we show that 
the axial and vector models are related by  canonical transformation 
(see \cite{dual} and refs. therein) 
 preserving the Hamiltonian which also   interchanges the topological and electric charges.

\section{General Construction of Toda Models}

The basic ingredient in constructing Toda models is the decomposition of a 
Lie algebra $\lie $ of finite  or infinite dimension in
terms of graded subspaces defined according to a grading operator $Q$, 
\br
\quad [ Q, \lie_l] = l \lie_l, \quad \quad 
\lie =\oplus \lie_l, \quad \quad [\lie_l, \lie_k ] \subset \lie_{l+k}, \; l,k= 0, \pm 1, \cdots 
\label{2.1}
\er
In particular, the zero grade subspace $\lie_0$ plays an important 
role since it is parametrized by the Toda fields.
The grading operator $Q$ induces the notion of negative and positive grade subalgebras
and henceforth the decomposition of a group element in the Gauss form, 
$ g=NBM$, 
where $N=\exp {(\lie_{<})}$, $B=\exp {(\lie_{0})}$ and $M=\exp {(\lie_{>})}$.
The action for the Toda fields is constructed from the gauged Wess-Zumino-Witten (WZW)
 action, 
\br
 & S_{G/H}(g,A,\bar{A}) = S_{WZW}(g) \nonumber \\
&- \frac{k}{2\pi}\int d^2x Tr( A(\bar{\partial}gg^{-1}-\epsilon_{+})
+\bar{A}(g^{-1}\partial g-\epsilon_{-})+Ag\bar{A}g^{-1}  ) 
\label{2.4}
\er
where $A = A_- \in \lie_{<}, \; \bar A = \bar A_+ \in \lie_{>}$, $\eps_{\pm}$ are 
constant operators of grade $\pm 1$.  The action (\ref{2.4}) is invariant under
\br
 g^{\prime}=\alpha_{-}g\alpha_{+}, \quad 
 A^{\prime}=\alpha_{-}A\alpha_{-}^{-1}
+\alpha_{-}\partial \alpha_{-}^{-1},
\quad  
 \bar{A}^{\prime}=\alpha_{+}^{-1}\bar{A}\alpha_{+}
+\bar{\partial}\alpha_{+}^{-1}\alpha_{+},  
\label{2.5} 
\er
 where $\a_{-}  \in \lie_{<}, \; \a_{+}  \in \lie_{>}$.
It therefore follows that  $ S_{G/H}(g,A,\bar{A}) =S_{G/H}(B,A^{\pr},\bar{A}^{\pr}) $.
 
Integrating over the auxiliary fields $A, \bar A$, we find the effective action,
\br
S_{eff}(B) =  S_{WZW}(B)- 
 {{k\o {2\pi}}} \int Tr ( \eps_+ B  \eps_- B^{-1}) d^2x   
  \label{2.6}
\er
The equations of motion are given by
\begin{eqnarray}
\bar \pa (B^{-1} \pa B) + [ {\eps_-}, B^{-1}  {\eps_+} B] =0, \quad  
\quad \pa (\bar \pa B B^{-1} ) - [ {\eps_+}, B {\eps_-} B^{-1}] =0
\label{2.7}
\er
It is straightforward to derive from the eqns. of motion (\ref{2.7}) that 
 chiral currents are associated to the subalgebra $\lie_0^0
\subset \lie_0$ defined as 
$\lie_0^0 = \{ X \in \lie_0, \;\;   [ X, \eps_{\pm} ] =0 \}$,
 i.e.,
\begin{eqnarray}
J_X = Tr ( X B^{-1} \pa B ), \quad \quad \bar J_X = Tr ( X \bar \pa B B^{-1}), 
\quad \quad \bar \pa  J_X = \pa \bar J_X =0
\label{2.7a}
\er
For the cases where $\lie_0^0 \neq 0$, we may impose consistently 
the additional constraints $J_{X} = \bar J_{X} = 0, X \in \lie_0^0$. 
The construction of the gauged WZW action  taking into account the
subsidiary constraints (\ref{2.7a}) reduces the model from the group $G_0$ to 
the coset $G_0/G_0^0$ and its action is given by
\br
S_{G_0/G_0^0}(B,{A}_{0},\bar{A}_{0} ) =  S_{WZW}(B)- 
 {{k\o {2\pi}}} \int Tr ( \eps_+ B  \eps_- B^{-1}) d^2x\nonumber \\   
  -{{k\o {2\pi}}}\int Tr( \pm  A_{0}\bar{\partial}B
B^{-1} + \bar{A}_{0}B^{-1}\partial B
\pm  A_{0}B\bar{A}_{0}B^{-1} + A_{0}\bar{A}_{0} )d^2x \nonumber \\
\label{2.12}
\er
where the $\pm $ signs correspond to axial or vector gaugings respectively.
The action (\ref{2.12}) is invariant under
\br
 B^{\pr} = \a_0 B \a_0^{\pr}, \quad \quad 
A_0^{\pr} = A_0 - \a_0^{-1} \pa \a_0, \quad \quad \bar A_0^{\pr} = \bar A_0 - \bar  \pa \a_0^{\pr}(\a_0^{\pr})^{-1}
\label{2.13}
\er
 where $\a_0^{\pr} =
\a_0(z, \bar z) \in \lie_0^0$ for axial and $\a_0^{\pr} =
\a_0^{-1}(z, \bar z) \in \lie_0^0$ for vector cases, i.e.,
$ S_{G_0/G_0^0}(B,{A}_{0},\bar{A}_{0} ) =  S_{G_0/G_0^0}(\a_0 B \a_0^{\pr} = g_0^f,{A^{\pr}}_{0},\bar{A^{\pr}}_{0} )$  

\section{The structure of the coset $G_0/G_0^0$}

We now discuss the structure of the coset $G_0/G_0^0$  constructed  
according to axial and vector gaugings.  We shall be considering  first the NA Toda 
models  where $\lie_0^0 = U(1)$.  The group element of 
the zero grade subgroup $G_0$ is parametrized as 
\br
B = e^{\tilde \chi E_{-\a_1}} e^{R\l_1 \cdot H + \sum_{l=1}^{n}\varphi_l h_l}e^{\tilde \psi E_{\a_1}}
\label{3.1}
\er
According to the axial gauging we can write $B$ as an element of the
the zero grade subgroup $G_0$ is parametrized as 
\br
B = e^{{1\o 2}R\l_1 \cdot H}( g_{0, ax}^f ) e^{{1\o 2}R\l_1 \cdot H}, \quad  
g_{0, ax}^f =  e^{\tilde \chi e^{{1\o 2}R}E_{-\a_1}} 
e^{ \sum_{l=2}^{n}\varphi_l h_l}e^{ \tilde \psi e^{{1\o 2}R} E_{\a_1}}
\label{3.2}
\er
The effective action is obtained  integrating (\ref{2.12}) 
over $A_0, \bar A_0$,  yielding \cite{tau}
\br
{\cal L}_{eff}^{ax}={1\o 2} \sum_{a,b =2}^{n}\eta_{ab} \pa \varphi_a \bar \pa \varphi_b  + 
{1\o 2} {{\bar \pa \psi \pa \chi }\o \Delta
 }e^{-\varphi_2}  -  V_{ax}, \quad \Delta = 1 + {{n+1}\o {2n}}\psi \chi
e^{-\varphi_2}
 \label{3.3}
 \er
 where $ \psi = \tilde \psi e^{{1\o 2}R}, \chi = \tilde \chi e^{{1\o 2}R}$, and 
  $V_{ax} = \sum_{l=2}^n e^{2\varphi_l - \varphi_{l-1}- \varphi_{l+1}} + 
 e^{\varphi_2+\varphi_n} ( 1+ \psi \chi
 e^{-\varphi_2})$.
 
The vector gauging can be implemented from the zero grade subgroup $G_0$ written as
\br
B = e^{u\l_1 \cdot H}( g_{0, vec}^f ) e^{-u\l_1 \cdot H}, \quad \quad {\rm where} \quad \quad
g_{0, vec}^f =  e^{\tilde \chi e^{u}E_{-\a_1}} 
e^{ \sum_{l=1}^{n}\phi_l h_l}e^{ \tilde \psi e^{-u} E_{\a_1}}
\label{3.3a}
\er
Since $u$ is arbitrary, we may choose $u = {1\o 2} ln ( {{\tilde \psi }\o {\tilde \chi }}) $ so that
\br
g_{0, vec}^f =  e^{t E_{-\a_1}} 
e^{  \sum_{l=1}^{n}\phi_l h_l}e^{t E_{\a_1}}, \quad \quad t^2 = \tilde \psi \tilde \chi 
\label{3.4}
\er
The effective action for the vector model is \cite{dual}   
\br
{\cal L}_{eff}^{vec}&=&{1\o 2} \sum_{a,b =1}^{n}\eta_{ab} \pa \phi_a \bar \pa \phi_b   
+ {{\pa \phi_1 \bar \pa \phi_1 }\o {t^2}}e^{\varphi_2 -2\phi_1} 
+ \pa \phi_1 \bar \pa ln (t) + \bar \pa \phi_1 \pa ln (t) - V_{vec}.
 \nonumber
 \er

We now discuss the simplest case in which $\lie_0^0$ is nonabelian, i.e. $\lie = \hat {SL}(3)$, $Q=d$, the homogeneous gradation 
and $\eps_{\pm} = \l_2 \cdot H^{(\pm 1)}$.  
In this case $\lie_0^0 =SL(2)\otimes U(1)$ is generated by $\lie_0^0 = \{ E_{\pm \a_1}, H_1, H_2\}$ 
and  $B$ is written as
\br
B &=& e^{\tilde \chi_1 E_{-\a_1}}e^{{1\o 2}(\l_1 \cdot H R_1 + \l_2 \cdot H R_2)} ( g_{0, ax}^f )
e^{{1\o 2}(\l_1 \cdot H R_1 + \l_2 \cdot H R_2)} e^{\tilde \psi_1 E_{\a_1}}\nonumber \\
g_{0, ax}^f &=& e^{\chi_1 E_{-\a_1-\a_2}+\chi_2 E_{-\a_2}} e^{\psi_1 E_{\a_1+\a_2}+\psi_2 E_{\a_2}}
\label{ax}
\er
where $\l_i, i=1,2$ are the fundamental weights of $SL(3)$.  The effective action is then obtained by 
integration over the auxiliary matrix fields  $A_0, \bar A_0$  yielding 
\begin{eqnarray}
{\cal L}_{eff}^{ax} = & {1\o {\Delta}}( {{\bar \pa \psi_2 \pa \chi_2 }} (1 + 
\psi_1\chi_1 + \psi_2 \chi_2 )  + 
{{\bar \pa \psi_1 \pa \chi_1} }(1 + \psi_2 \chi_2 ) )  \nonumber\\
&  - {1\o {2\Delta}}( \psi_2 \chi_1 {{\bar \pa \psi_1 \pa \chi_2 }}+ \chi_2 \psi_1 {{\bar \pa \psi_2 \pa \chi_1 }}) -V 
\label{6.6}
\er
where $V =  {2\o 3} + \psi_1 \chi_1 + \psi_2 \chi_2 $ and 
$\Delta = (1+ \psi_2 \chi_2 )^2 + \psi_1 \chi_1 (1 + {3 \o 4} \psi_2 \chi_2 )$.
For the vector action, the zero grade group element $B$ in (\ref{ax}) is parametrized as 
\begin{eqnarray}
B &=&e^{\tilde \chi_1 E_{-\a_1}}e^{{1\o 2}(\l_1 \cdot H u_1 + \l_2 \cdot H u_2)} ( g_{0, vec}^f )
e^{-{1\o 2}(\l_1 \cdot H u_1 + \l_2 \cdot H u_2)}
e^{\tilde \psi_1 E_{\a_1}}
\label{vec}
\er
where 
$g_{0,vec}^f = e^{-t_2 E_{-\a_2}- t_1 E_{-\a_1-\a_2}} 
e^{\phi_1 h_1 + \phi_2 h_2}e^{t_2 E_{\a_2}+ t_1 E_{\a_1+\a_2}}$.
The effective action is then \cite{spin}
\br
{\cal L}_{vec} &=& {1\o 2}\sum_{i=1}^{2} \eta_{ij}\pa \phi_i \bar \pa \phi_j +
 {{\pa \phi_1 \bar \pa \phi_1}\o {t_1^2}}e^{-\phi_1-\phi_2} +\bar \pa \phi_1 \pa ln (t_1) +  
 \pa \phi_1 \bar \pa ln (t_1)  \nonumber \\
 &-&
 {\pa \phi_1 \bar \pa \phi_1}{( {{t_2}\o {t_1}})^2}e^{-2\phi_1 +\phi_2}  
 + {{\bar \pa (\phi_2 -\phi_1)\pa (\phi_2 - \phi_1)}\o {t_2^2}}e^{\phi_1-2\phi_2}\nonumber \\
   &+& \bar \pa (\phi_2 - \phi_1) \pa ln (t_2) +  
 \pa (\phi_2- \phi_1) \bar \pa ln (t_2)  -V 
 \label{svec}
 \er
 where $V = {2\o 3}  - t_2^2 e^{-\phi_1 +2 \phi_2} - t_1^2 e^{\phi_1+\phi_2}$ and 
 $\eta_{ij} = 2 \d_{ij}-\d_{i, j-1}-\d_{i, j+1}$.

\section{Axial-Vector Duality}

In this section we shall prove that the axial and vector models are 
related by a canonical transformation.
Consider the $SL(3)$  vector model 
\br
{\cal L}_{vec} &=& \pa \phi_1 \bar \pa \phi_1 + \pa \phi_2 \bar \pa \phi_2 -{1\o 2} \pa \phi_2 \bar \pa \phi_1-
{1\o 2} \pa \phi_1 \bar \pa \phi_2 
+ {{\pa \phi_1 \bar \pa \phi_1 }\o {t^2}}e^{\varphi_2 -2\phi_1} \nonumber \\
&+& \pa \phi_1 \bar \pa ln (t) + \bar \pa \phi_1 \pa ln (t) - (e^{\phi_1-2\phi_2} +e^{-\phi_1+2\phi_2}-t^2e^{\phi_1+\phi_2}).
 \label{5.1}
\er
  In terms of  the new set of more convenient variables 
$a= (1-t^2e^{2\phi_1-\phi_2})  $, 
$\; \; f= \phi_1-2 \phi_2 $, $ \theta = \phi_1$
the lagrangian (\ref{5.1}) becomes
\br
4{\cal L}_{vec} =({{1+3a}\o {1-a}}) \pa \theta \bar \pa \theta  
- \pa \theta (\bar \pa f + {{2\bar \pa a}\o {1-a}}) 
- \bar \pa \theta  (\pa f + {{2\pa a}\o {1-a}}) + \pa f \bar \pa f - 4V_{vec} 
\label{5.3} 
\er
where $V_{vec} = e^f + {{a}}e^{-f}$.  The canonical momenta are given by $ \Pi_{\rho} = 
{{\d {\cal L}_{vec}}\o {\d \dot \rho}}$, $\rho = \theta, f,
a$.   The hamiltonian is then given by
\br
{\cal H}_{vec} &=& -(1-a)\Pi_{a} \Pi_{\theta} + \Pi_f^2 - (1-a)\Pi_a \Pi_f - a(1-a) \Pi_a^2 \nonumber \\
&+& {1\o 4} {{(1+3a)}\o {1-a}} {{\theta}^{\pr}}^2 -{1\o 2}( f^{\pr} +{{2a^{\pr}}\o {1-a}}){\theta}^{\pr} - {1\o 4}
{f^{\pr}}^2  + V_{vec}
\label{5.5}
\er
Consider now the following modified lagrangian 
\br
{\cal L}_{mod} ={\cal L}_{vec} - \tilde \theta (\pa \bar P - \bar \pa P )
\label{5.6}
\er
where we identify $\pa \theta = P, \quad \bar \pa \theta = \bar P$ \cite{buscher}.  
Integrating by parts,
\br
{\cal L}_{mod} =({{1+3a}\o {1-a}})  P \bar P  
-{1\o 4} P  (\bar \pa f + {{2\bar \pa a}\o {1-a}} + \bar \pa \tilde \theta ) \nonumber \\
-{1\o 4} \bar P   ( \pa f + {{2\pa a}\o {1-a}} - \pa \tilde \theta ) + {1\o 4}\pa f \bar \pa f - V_{vec} 
\label{5.7}
\er
Integrating over the auxiliary fields $P$ and $\bar P$ we find the effective action
\br
{\cal L}_{eff} ={1\o 4}\pa f \bar \pa f - 
{1\o 4}{{(1-a)}\o {1+3a}} (\bar \pa f + {{2\bar \pa a}\o {1-a}} + \bar \pa \tilde \theta ) 
( \pa f + {{2 \pa a}\o {1-a}} -  \pa \tilde \theta ) -V 
\label{5.8}
\er
with   canonical momenta defined by $\; \Pi_{\rho} = 
{{\d {\cal L}_{eff}}\o {\d \dot \rho}}$, $ \; \rho = \tilde \theta, f,
a$. The hamiltonian  becomes
\br
{\cal H}_{mod} &=& -{1\o 2}(1-a)\Pi_{a} {\tilde \theta}^{\pr} + \Pi_f^2 - (1-a)\Pi_a \Pi_f - a(1-a) \Pi_a^2 \nonumber \\
&+&  {{(1+3a)}\o {1-a}} {\Pi_{\tilde \theta}}^2 -{1\o 2}( f^{\pr} +{{2a^{\pr}}\o {1-a}})\Pi_{\tilde \theta} - {1\o 4}
{f^{\pr}}^2  + V_{vec}
\label{5.10}
\er
The canonical transformation 
\br
\Pi_{\theta} = -{1\o 2} \tilde \theta^{\pr}, \quad \quad \theta ^{\pr} = -2 \Pi_{\tilde \theta}
\label{5.11}
\er
preserves the Poisson bracket structure and provide the 
 equality of the hamiltanians ${\cal H}_{mod} = {\cal H}_{vec}$. 
If we now substitute
\br
\tilde \theta = 2 ln ( {{\psi}\o {\chi}}), \quad a = 1+ \psi \chi e^{-\varphi_2}, \quad f = - \varphi_2
\label{5.12}
\er
in the effective  lagrangian $ {\cal L}_{eff}$ (\ref{5.8}), we find
\br
{\cal L}_{eff} = {1\o 2}\pa \varphi_2 \bar \pa \varphi_2 + {{\pa \chi \bar \pa \psi }\o {\Delta }}e^{-\varphi_2} - V
\label{5.13}
\er
which is precisely the axial lagrangian (\ref{3.3}) for $\lie_0 = SL(3)$.  It therefore  becomes clear that 
the axial and the vector models are related by  the
canonical transformation (\ref{5.11}) which preserves their hamiltonians.

For the case of $\lie_0^0 =SL(2)\otimes U(1)$ we found the  canonical transformation 
 responsible for the equality of the Hamiltonians 
 to be \cite{spin}
\br
 \Pi_{\theta_{\a}} = -2\pa_x \tilde \theta_{\a}, \quad \quad  \Pi_{\tilde \theta_{\a}} = -2\pa_x \theta_{\a}, \quad \a =1,2
\label{cc}
\er
where 
$ \theta_{\a} = ln ( {{\psi_\a} / {\chi_\a}} ), \quad \tilde \theta_1 = -{1\o 2}\phi_1, \quad 
\tilde \theta_1 = -{1\o 2}( \phi_1 + \phi_2)$

As a last comment of this section we should like to analyse the topological and Noether charges 
of the axial and vector models.  Consider the  first example where $\lie_0^0 = U(1)$ and with   vector and axial lagrangians
 given by (\ref{5.3}) and (\ref{5.8}) respectively. The Noether charges associated to the global transformation 
$\theta \rightarrow \theta + c$ and $\tilde \theta \rightarrow \tilde \theta + \tilde c$ where $ c, \tilde c = $ constant,
 are given by
\br
Q^{Noether}_{vec} &=& \int ( {{\d {\cal L}_{vec}}\o {\d \dot \theta}})\d \theta dx  
=\int  \Pi_{\theta} dx, \nonumber \\
Q^{Noether}_{ax} &=& \int ( {{\d {\cal L}_{ax}}\o {\d \dot {\tilde \theta}}})\d \tilde \theta dx 
= \int  \Pi_{\tilde \theta} dx
\label{noether}
\er
Since the vector and axial  models possess respectively the foloowing  topological charges 
\br
Q^{Top}_{vec} = \int (\pa_x \theta ) dx , \quad \quad Q^{Top}_{ax} = \int (\pa_x \tilde \theta ) dx
\label{top}
\er
it is clear  that under the canonical transformation (\ref{5.11}) 
their Noether and topological charges become   interchanged.
The same can be extended to all isometric variables within the models described by 
lagrangians (\ref{6.6}) and (\ref{svec}).

\section{Concluding Remarks}

We have seen that the crucial ingredient which allows the construction of the axial 
and vector models is the existence of a non trivial subgroup $G_0^0$. We have worked out 
explicitly examples in which $G_0^0 = U(1)$ and $G_0^0 = SL(2)\otimes U(1)$ 
involving one and two isometric variables  $\theta_{\a}$.  
The same strategy works equally well for  generalized  multicharged NA Toda models.

An interesting and intriguing subclass of NA Toda models correspond to 
the following three affine  Kac-Moody algebras, 
$B_n^{(1)}, A_{2n}^{(2)}$ and $D_{n+1}^{(2)}$.  Their axial and vector
actions were constructed in \cite{dual} and shown to be identical.  
In fact, those affine algebras satisfy the {\it no torsion condition} 
proposed in \cite{dual} which is fulfilled by  Lie algebras possesseing 
$B_n$-tail like Dynkin diagrams.
 The very same  selfdual models were shown to possess an exact S-matrix  
 coinciding with certain Thirring models coupled to
 affine abelian Toda models in ref. \cite{Fat}.

\bigskip
{\small  We are grateful to  CNPq, FAPESP and UNESP for 
financial support.}


\bigskip

\bbib{9}               
\bibitem{tau} J.F. Gomes, E.P. Gueuvoghlanian, G.M. Sotkov and A.H.
Zimerman, {{\sl Nucl. Phys.} {\bf B598} (2001) 615}, hepth/0011187; 
{{\sl Nucl. Phys.} {\bf B606} (2001) 441}, hepth/0007169




\bibitem{dual} J.F. Gomes, E.P. Gueuvoghlanian, G.M. Sotkov and A.H.
Zimerman, {{\sl Ann. of Phys.} {\bf 289} (2001) 232}, hepth/0007116



\bibitem{spin}J.F. Gomes,G.M. Sotkov and A.H.
Zimerman, {{\sl J. Physics} {\bf A37} (2004) 4629}


\bibitem{buscher}T. Busher, {{\sl Phys. Lett.} {\bf 159B} (1985) 127}, {{\sl Phys. Lett.} {\bf 194B} (1987) 59},
{{\sl Phys. Lett.}} {\bf 201B} (1988) {466}

\bibitem{Fat}  V.A. Fateev, {{\sl Nucl. Phys.}  {\bf B479} (1996) 594}
 


\ebib
\end{document}